\documentclass[12pt]{article}
\usepackage[utf8]{inputenc}
\usepackage{amsmath}
\usepackage{amsfonts}
\usepackage{amssymb}
\usepackage{epsfig}
\usepackage{epstopdf}
\usepackage{indentfirst}
\usepackage{geometry}
\usepackage{graphicx}
\usepackage{caption}
\usepackage{subfig}
\usepackage[countmax]{subfloat}
\usepackage{rotating}
\usepackage{hyperref}
\usepackage{lscape}
\usepackage[numbers,sort&compress]{natbib}
\usepackage{tabularx}
\usepackage{setspace}
\usepackage{enumerate}
\usepackage{lineno}
\usepackage[section]{placeins}

\newtheorem{theorem}{Theorem}[section]

\newtheorem{definition}{Definition}[section]

\numberwithin{equation}{section}
\def\bbR{{\mathbb R}}

\def\bA{{\boldsymbol A}}
\def\bB{{\boldsymbol B}}
\def\bC{{\boldsymbol C}}

\def\bH{{\boldsymbol H}}

\def\bO{{\boldsymbol O}}

\def\bC{{\boldsymbol C}}

\def\bS{{\boldsymbol S}}

\def\b1{{\boldsymbol 1}}
\def\b0{{\boldsymbol 0}}

\def\bbC{{\mathbb C}}

\def\bbI{{\mathbb I}}

\def\cH{{\mathcal H}}

\def\cR{{\mathcal R}}
\def\cS{{\mathcal S}}

\def\cE{{\mathcal E}}

\graphicspath{%
    {converted_graphics/}
    {/}
}

\begin{document}

\title{On the EPR paradox in systems \\with finite number of levels (Revised)}
\author{Henryk Gzyl \\
Center for Finance, IESA School of Business\\
{ henryk.gzyl@iesa.edu.ve }
} 

\date{}
 \maketitle

\baselineskip=1.5\baselineskip \setlength{\textwidth}{6in}

\begin{abstract}
In this work we reexamine the EPR paradox for composite systems with a finite number of levels. The analysis emphasizes the connection between measurements and conditional probabilities. This connection implies  that when a measurement is performed, the microscopic states compatible with the measurement is different from the class of all possible microscopic states, therefore the new quantum state and the probability distribution change and become a function of the observable being measured. Therefore, the predictions that one can make given the knowledge of the result of a measurement change.

Systems with finitely many levels are simpler to describe because the analysis is not encumbered by the mathematical technicalities of the continuous case, the underlying physical interpretations are the same  and the experimental setups used to test quantum mechanics with the paradox in mind finitely many levels.
\end{abstract}


\begin{spacing}{0.1}
   \tableofcontents
\end{spacing}

\section{Introduction and preliminaries} 
It was 90 years ago that Einstein, Podolsky and Rosen published a paper, \cite{EPR}. in which they examined the basic understanding of the probabilistic interpretation of quantum mechanics, of its philosophical interpretation, of what is reality, and what it means to measure and to predict.

In that work they presented an example that seemed to contradict the foundations of quantum mechanics. The essence of the argument goes as follows. They consider a system made up of two  particles. Suppose that there are no external forces on the system and that the total momentum is conserved. They consider the system to be in an entangled total state of known total momentum. Then they argue that if one measures the momentum of one of the particles, since the total momentum is given, the momentum of the other is known with certainty. But then they claim that this contradicts the uncertainty principle, because that means that one can measure the position of the second particle and record the variance of the measurements, which will be finite and violate the uncertainty principle, since its momentum is known with certainty.   

The EPR paradox in quantum systems with a finite number of levels and its standing in relation to other theoretical issues in quantum mechanics, from the theoretical and experimental point of view is discussed in many places . See for example \cite{A}, \cite{La} and \cite{LeB}. An important collection, organized by topic and by year is posted at \cite{www} .
 
 In order to rephrase the argument of EPR for quantum systems with a finite number of levels, we need to establish some notations and terminology. We consider a system made up of two identical components. To describe any of them, let us denote by $\cH_0$ (or by $\bbC^N$) the $N$-dimensional complex Hilbert space in which the physical states are the unit vectors of $\cH_0.$ States in the Hilbert space denoted by the standard $\Psi$ or $\phi,$ and we sometimes we use the alternative standard Dirac notation $|\Psi\rangle$ of $|\phi\rangle.$ 

We suppose that there exist an observable, or Hermitian matrix  $\bA$ that has $N$ distinct eigenvalues $\cR_A=\{a_1<a_2<...<a_N\}$ whose eigenvectors $\{|a_n\rangle: n=1,...,N\}$ span $\cH_0.$  The quantities $p_\Psi(a_i) = \langle\Psi|a_i\rangle\langle a_i|\Psi\rangle|^2$ are interpreted as the probability of observing the value $a_i$ when measuring the observable $\bA$ when the system is in the state $\Psi.$

If $f:\{a_1,...,a_N\}\to \bbR$ is any function whose values depend on the observed value of $\bA,$ the best predictor of $f(\bA)$ is the following

\begin{equation}\label{BP1}
\cE_\Psi[f(\bA)] = \sum_n f(a_n)p_\Psi(a_i) = \langle\Psi|a_i\rangle\langle a_i|\Psi\rangle|^2 = \langle\Psi|f(\bA)|\Psi\rangle.
\end{equation}

The middle term in the first line of \eqref{BP1} is called the best predictor of $f(\bA)$ because it  minimizes the prediction error defined by $\sum_n \big(f(a_n)-m\big)^2p_\Psi(a_i).$  The smallest possible error is:
\begin{equation}\label{BP2}
\cE_\Psi[\big(f(\bA)-\langle\Psi|f(\bA)|\Psi\rangle)^2]= \big(\Delta_\Psi(\bA)\big)^2.
\end{equation}

We assume furthermore, that there exist matrices  $\bB,\bC$ such that the following (commutation) relationship exists between the operators $\bA,\bB,\bC.$

\begin{equation}\label{com1}
[\bA,\bB] = i\alpha\bC.
\end{equation}
(A typical example is provided by the Pauli spin matrices). The constant $\alpha$ depends on the system of units. We suppose that $\alpha=1.$ An interesting consequence of this identity is contained in:
\begin{theorem}\label{T1}
Suppose that the triplet $\bA,\bB,\bC$ of Hermitian matrices satisfy \eqref{com1}. Let $\phi_a$ be such that $\bA\phi_a=a\phi_1.$ Then:
\begin{gather}
tr(\bC)=0.\;\;\;\label{T1.1}\\
\langle\phi_a,\bC\phi_a\rangle=0\;\;\;\label{T1.2}
\end{gather}
In particular, the trace $tr{\bC}=0.$
\end{theorem}
The proof is simple and left for the reader. Identity \eqref{T1.2} will be relevant for the analysis of the uncertainty principle for predicting when there is no error in the prediction.

A foundational stone of quantum mechanics is the Heisenberg uncertainty principle, which states that for any three observables $\bA, \bB, \bC$ related by \eqref{com1}, the product of the prediction (or measurement) errors of $\bA$ and $\bB$ is bounded below as follows:

\begin{equation}\label{HUP}
\Delta_\Psi(\bA)\Delta_\Psi(\bB) \geq \frac{1}{2}|\langle\Psi|\bC|\Psi\rangle|.
\end{equation}

When studying systems with finitely many levels and $\Psi$ is such that $\langle\Psi|\bC|\Psi\rangle=0,$ then in this case one or both of the observables can be predicted with zero error without violating \eqref{HUP} .

The state space of the joint systems is $\cH=\cH_0\otimes\cH_0.$ Now we consider two copies on the operators  $\bA,\bB,\bC,$ denoted by  $\bA_i,\bB_i,\bC_i$ with $i=1,2$ when we refer to the operators as being observables of the first $i=1$ or respectively  $i=2$ of the second component. We emphasize that 
actually, $\bA_1$ is shorthand for $A_1\otimes\bbI,$ where $\bbI$ denotes the identity operator on $\cH_0.$  Similarly  $\bA_2$ is shorthand for $\bbI\otimes\bA_2.$ We use the formal notation when clarity requires it.

We furthermore suppose that the Hamiltonian operator $\bH$  that describes the time evolution of the joint system is such that  $[\bH, \bA_1+\bA_2]=0.$ That is the sum $\bS=\bA_1+\bA_2$ is a conserved quantity.\\
\textbf{Statement of the EPR paradox for systems with finitely many levels}\\
Suppose that the joint system has been prepared in a state $\Psi.$ Suppose furthermore that $\bS$ is measured and the value $s$ is obtained.  Suppose now that we measure $\bA_1$ of subsystem 1 and a value $a_1$ is obtained. Then EPR rightly asserted that, with probability $1,$ if we were to observe $\bA_2$ we would obtain $a_2=s-a_1.$ Nothing wrong nor paradoxical with that. Then they argued as follows: Since we know that $\bA_2=a_2$ with probability $1$ without measuring,  and therefore with $0$ error, then we could measure $\bB_2$ with any precision, and that would contradict the uncertainty relationship \eqref{HUP}.

It is the aim of this work to explain why there is no paradox at all. For that we need the relationship between prediction with respect to post measurement states and quantum conditional probabilities. What happens is that after measurement the probability distribution of the values of the observables changes.  At this point we mention \cite{He} and \cite{Ho} for the use of probabilistic and statistical tools for prediction in quantum signal processing, and \cite{J} for the basics of quantum measurement theory. For the relationship between classical and quantum probability, the reader may consider \cite{Gu1} and \cite{Gu2}. The relationship between these themes the way we use it in this work is established in Section \ref{JS}.

After that lengthy notational preamble, let us describe the contents of the paper and its relationship to the EPR paradox. In Section \ref{JS} we describe the joint system and how to prepare a state of given value of the sum of the two basic observables. If we denote by $\bS=\bA_1+\bA_2$ the total ``$\bA$'' operator, we will prove that given that a value  $s$ of $\bS$ is observed, the probability distribution of the values of  $\bA_1,$ say, is the obtained by conditioning the joint probability distribution of the pair $(\bA_1,\bA_2).$ 

In Section \ref{EPR} we examine how  the change in the probability distribution bears on the analysis of the EPR problem for a system with a finite number of levels. We consider two cases. First, we examine the predictions and the prediction errors of the basic observables of each subsystem if the state is a state of known value of $\bS.$ After that we reexamine the same predictions if we suppose that we first measure $\bA_1$ to obtain a state of given value of $\bA_1.$ 

In section 4 we repeat the same analysis when $\bA,\bB,\bC$ are the Pauli spin matrices and each subsystem is a two level system. We and with a short summary of the main ideas developed in the paper. 

In Section 5 we recall the notion of quantum conditional expectation and verify that predicting in the post-measurement state coincides with the quantum conditional expectation. Therefore, consistency is maintained. We close with a few remarks.


\section{Quantum measurement and conditional probabilities}\label{JS}
To analyze the composite system we label by $\cH_1$ and $\cH_2$ two copies of the state space, and form the (tensor) product $\cH=\cH_0\otimes\cH_0.$ We add a subscript $i=1,2$ to the symbols introduced above to denote the operators acting on the corresponding factors, and by $\{|a_n(1),a_m(2)\rangle = |a_n(1)\rangle\otimes|a_m(2)\rangle: 1\leq n,m\leq N\}$ the basis vectors in the product state. If the two components are identical, the notation $\{|a_n,a_m\rangle$ is less cumbersome.  We use $\bA_1$ instead of the formal $\bA_1\otimes\bbI$ to mean $\bA_1|a_n(1),a_m(2)\rangle = \big(\bA_1|a_n(1)\rangle\big)\otimes|a_m(2)\rangle).$

Consider now the sum $\bS=\bA_1+\bA_2,$ that is, the total $\bA$ of the system. The range of this operator is $\cR_S=\{a_n(1)+a_m(2): (a_n(1),a_m(2))\in\cR_A\times\cR_A\}.$ Let us denote the possible eigenvalues of $\bS$ by $\{s_1<s_2<...<s_D.\}$ The minimum being $s_1=a_1(1)+a_1(2)$ and the maximum $s_D=a_N(1)+a_N(2).$ And let us put $D$ for the number of different eigenvalues. Let us denote by $\cS(k)$ the anti-diagonals in $\cR_A\times\cR_A$ along which the sum $a_n(1)+a_m(2)$ is constant, that is:
$$\cS(k)=\{(a_n(1),a_m(2))\in\cR_A\times\cR_A: a_n(1)+a_m(2)=s_k;\;1\leq k\leq D.\}$$ 
Let us write $C(k)$ for the cardinality of $\cS(k).$ This means that $C(k)$ is the degeneracy of the eigenvalue $s_k,$ since for $(a_n(1),a_m(2))\in\cS(k)$ we have
$\bS|a_n(1),a_m(2)\rangle=s_k|a_n(1),a_m(2)\rangle.$ We have enough notations to write down the probability $p_\Psi(s_k)$ of observing the value $s_k$ in an experiment. The elementary way, which start from identifying $q_\Psi(n,m)=|\langle a_n(1),a_m(2)|\Psi\rangle|^2$ as the probability that system in the quantum state $\Psi$ yields values $(a_n(1),a_m(2))$ for the observables $\bA_1,\bA_2.$ Since observing $s_k$ is equivalent to observing any of the vales $(a_n(1),a_m(2))$ in $\cS_(k),$ it follows that:
$$p_\Psi(s_k)=\sum_{(n,m)\in\cS(k)}q_\Psi(n,m).$$
Introducing the logical counting function $\chi(a_n,s_k),$ that assumes value $1$ if there exists an $a_m$ such that $a_n+a_m=s_k,$ and assumes value $0$ otherwise, the previous identity can be written:  

\begin{equation}\label{ps1}
p(s_k)=\sum_{(n,m)\in\cS(k)}q_\Psi(n,m)= \sum_{n=1}^N |\langle a_n(1),s_k-a_n(1)|\Psi\rangle|^2\chi(a_n(1),s_k).
\end{equation}
The projection $\Pi_k$ onto the subspace spanned by the eigenvalues of $\bS$ corresponding to $s_k,$ that is defined as:

\begin{equation}\label{proj}
\Pi_k = \sum_{(n.m)\in\cS(k)}|a_n(1),a_m(2)\rangle\langle a_n(1),a_m(2)|=\sum_{n=1}^N|a_n,s_k-a_n\rangle\langle a_n,s_k-a_n|\chi(a_n,s_k).
\end{equation}
It is easy to verify that: 
\begin{equation}\label{tr}
tr\big(\Pi_k|\Psi\rangle\langle\Psi|\big) = \sum_{n=1}^N|\langle a_n(1),s_k-a_n(1)|\Psi\rangle|^2\chi(a_n,s_k) = p(s_k).
\end{equation}
With these notations we can write the decomposition of the state $\Psi$ in terms of eigenstates of $\bS$ as follows.

\begin{equation}\label{decomp}
\begin{aligned}
&|\Psi\rangle = \sum_{(n,m)}|a_n(1),a_m(2)\rangle\langle a_n(1),a_m(2)|\Psi\rangle\\
& = 
\sum_{k=1}^D\sum_{n=1}^N|a_n,s_k-a_n\rangle\langle a_n,s_k-a_n|\Psi\rangle\chi(a_n,s_k)\\
& = \sum_{k=1}^D\bigg(\frac{\Pi_k|\Psi\rangle}{\sqrt{p(s_k)}}\bigg)\sqrt{p(s_k)} = \sum_{k=1}^D \sqrt{p(s_k)}|\Psi_{s_k}\rangle.
\end{aligned}
\end{equation}
Here we introduced the notation:
\begin{equation}\label{eigv}
|\Psi_{s_k}\rangle = \frac{\Pi_k|\Psi\rangle}{\sqrt{p(s_k)}}.
\end{equation}
This is the normalized orthogonal projection of $\Psi$ onto the eigenspace of $\bS$ corresponding to the eigenvalue $s_k.$ Note that $\Psi_{s_k}$ is normalized, and it is a weighted linear superposition of the vectors spanning the subspace corresponding to the eigenvalue $s_k.$ It is interesting to add that $\Psi_{s_k}$  happens to be an entangled state. This will be the starting point of the lack of paradox in the EPR analysis.

To bring forth the analogy with quantum conditional probability (see Section 5), and to think of the quantum state after measurement, and of the conditional probabilities as functions of the observable being measured, we may write:
\begin{equation}\label{rs1}
\Psi_{\bS}= \frac{\Pi_{\bS}|\Psi\rangle}{\sqrt{p(\bS)}}.
\end{equation}
where both, the projection operator $\Pi_{\bS}$ and the probability $p(\bS)$ are functions of $\bS$ that assume values, respectively, $\Psi_{s_k}$ and $p(s_k)$ when $\bS=s_k.$ 
\section{Prediction post measurements and the EPR paradox}\label{EPR}
We suppose that $\bS$ commutes with the $\bH,$ that is, the observed value of $\bS,$ remains constant during the time evolution of the system. 

  Suppose as well, that we measure the value of the observable $\bS$ and obtain  an eigenvector $|\Psi_s\rangle$ in which $\bS$ has value $s$ as specified in \eqref{eigv}.

There are two different questions that we can ask ourselves: First, what are the predicted values and the prediction errors of the observables $\bA_i$ and $\bB_i$ for $i=1,2.$  Second, we can ask the same question, when the state is the post-measurement state after measuring, $\bA_1$ say.

\subsection{Prediction given a prior measurement of $\bS$}  The question is: What is the expected value of a function $f(\bA_1)$ of the observable $\bA_1$ in the stated $|\Psi_s\rangle$? In the notations of \eqref{ps1} the answer is:
\begin{equation}\label{pred3}
\langle\Psi_s,f(\bA_1)\Psi_s\rangle=  \sum \frac{\langle\Psi|,\Pi_s f(\bA_1)\Pi_s|\Psi\rangle}{p(s)} = \sum_{n=1}^N f(a_n)\frac{q_{\Psi}(a_n,s-a_n)}{p(s)}\chi(a_n,S(s)).
\end{equation}
If we think of $q_\Psi(a_n,a_m)$ as the probability of observing $a_n$ and $a_m$ respectively when measuring $\bA_1$ and $\bA_2,$ then
$$\frac{q_{\Psi}(a_n,s-a_n)}{p(s)}$$
is the conditional probability of observing the value $a_n$ when measuring $\bA_1$ given that $\bS=\bA_1+\bA_2$ has been observed to have value $s.$

To stress the fact that there is an underlying dependence on the observable $\bS,$ we make use of the notations introduced in \eqref{rs1} and rewrite \eqref{pred3} as:
\begin{equation}\label{rs2}
\langle\Psi_{\bS},f(\bA_1)\Psi_{\bS}\rangle=  \sum \frac{\langle\Psi|,\Pi_{\bS} f(\bA_1)\Pi_{\bS}|\Psi\rangle}{p(\bS)} = \sum_{n=1}^N f(a_n)\frac{q_{\Psi}(a_n,\bS-a_n)}{p(\bS)}\chi(a_n,\bS).
\end{equation}

To continue, applying \eqref{pred3} to $f(\bA_1)=\bA_1$ and $f(\bA_1)=\bA_1^2,$ we obtain:
\begin{gather}
\langle\Psi_s,\bA_1\Psi_s\rangle = m(A_1) = \sum_{n=1}^N f(a_n)\frac{q_{\Psi}(a_n,s-a_n)}{p(s)}\chi(a_n,S(s))\;\;\label{mean}\\
\sqrt{(\langle\Psi_s,(\bA_1-m(\bA_1))^2\Psi_s\rangle}=\Delta_\Psi(A_1)=\sqrt{\sum_{n=1}^N (a_n)^2\frac{q_{\Psi}(a_n,s-a_n)}{p(s)}\chi(a_n,S(s))-m(\bA_1)^2} \;\;\label{stdv}
\end{gather}
In the standard (non-quantum) probability theory, \eqref{pred3} is called the conditional expected value of $f(\bA_1)$ given that $\bS=s.$ In this case, one thinks of \eqref{pred3} as the realized value of a random variable that depends on $\bS.$

The previous computations can be carried out for $\bA_2.$ Using the fact that $\bS$ is given, the next statements are intuitive.
\begin{theorem}\label{T2}
Suppose $\bS$ is observed to be $s$ and the system is left in the state $\Psi_s.$ then
\begin{gather}
\langle\Psi_s,\bA_2\Psi_s\rangle = s-\langle\Psi_s,\bA_1\Psi_s\rangle.\;\;\;\label{equi1}\\
\Delta_\Psi(A_1) = \Delta_\Psi(A_2).\;\;\;\label{equi2}
\end{gather}
\end{theorem}
To verify the first assertion note that:
$$m(\bA_2)=\langle\Psi_s,\bA_2\Psi_s\rangle=\langle\Psi_s,(\bS-\bA_1)\Psi_s\rangle=s-m(\bA_1).$$
The intermediate steps in the following computation also follow from the fact that $\Psi_s$ is an eigenvector of eigenvalue $s,$ the commutativity between $\bS$ and $\bA_1)$ and the previous computation.
$$\begin{aligned}
\langle\Psi_s,(\bA_2-m(\bA_2))^2\Psi_s\rangle = \langle\Psi_s,(\bS-s-(\bA_1-m(\bA_1))^2\Psi_s\rangle\\
\langle\Psi_s,(\bS-s)^2\Psi_s\rangle-2\langle\Psi_s,(\bA_-m(\bA_1))(\bS-s)\Psi_s\rangle+\langle\Psi_s,(\bA_1-m(\bA_1))^2\Psi_s\rangle.
\end{aligned}$$
The first and second term vanish. Thus the result is \eqref{equi2}. 

This theorem asserts that, given that their sum $\bS$ is constant, the variability of $\bA_1$ compensates that of $\bA_2,$ so that the errors in predicting one of them, equal that of predicting the other in the post-measurement state.

Not only that, the probabilities of observing $a_2=s-a_1$ in the stated $\Psi_s$ coincide. Both equal $q_\Psi(a_1,s-a_1)/p_\Psi(s).$ That is, observing $\bA_1$ yields complete information about the possible outcomes of $\bA_2.$ No contradiction to the uncertainty principle occurs at this level.

\subsection{Prediction after measuring $\bS$ and then $\bA_1$}
For a system in the state $\Psi_s,$ a measurement of $\bA_1$ yields a value $a_i(1)$ with probability $q_\Psi(a_i(1),s-a_i(1))/p(s),$ and leaves the system in a state
\begin{equation}\label{red1}
\phi_{a_i(1)}=\frac{\Pi_{a_i(1)}\Psi_s}{tr(\Pi_{a_i(1)}|\Psi_s\rangle\langle\Psi|)}.
\end{equation}
Since $\Pi_{a_i(1)}=|a_i(1)\rangle\langle a_i(1)|\otimes\bbI,$ then 
$$\Pi_{a_i(1)}\Psi_s=\frac{\langle a_i(1),s-a_i(1)|\Psi\rangle}{p(s)}|a_i(1),s-a_i(1)\rangle.$$
and
$$tr(\Pi_{a_i(1)}|\Psi_s\rangle\langle\Psi|)=\frac{q_\Psi(a_i(1),s-a_i(1))}{p(s)}.$$
Therefore, \eqref{red1} written out explicitly is
\begin{equation}\label{red2}
\phi_{a_i(1)}=\frac{\langle a_i(1),s-a_i(1)|\Psi\rangle}{|\langle a_i(1),s-a_i(1)|\Psi\rangle|}|a_i(1),s-a_i(1)\rangle.
\end{equation}
Therefore, the expected value of a function $g(\bA_2)$ in this state is
\begin{equation}\label{rev1}
\langle\phi_{a_i(1)},g(\bA_2)\phi_{a_i(1)}\rangle=\langle\phi_{a_i(1)},g(\bS-\bA_1)\phi_{a_i(1)}\rangle=g(s-a_i(1)).
\end{equation}
In particular, the probability of measuring the value $a_j(2)$ of $\bA_2$ in this state is
\begin{equation}\label{red3}
|\langle a_i(1),a_j(2)|\phi_{a_i(1)}\rangle|^2 = \delta_{a_j(2),s-a_i(1)}.
\end{equation}
This is a compact way of saying that the probability is $1$ if $a_j(2)=s-a_i(1)$ or $0$ otherwise.

 In the operator valued notations, we restate two variations of the theme of \eqref{mean} and\eqref{stdv} as:
\begin{gather}
\langle\phi_{a_i(1)},\bA_2\phi_{a_i(1)}\rangle = \bS-\bA_1,\;\;\;\label{mean.1}\\
\Delta_{\phi_{a_i(1)}}(\bA_2) = 0.\;\;\;\label{stdv.2}
\end{gather}
The second assertion is clear. If in the state $\phi_{a_1}$ we predict the value of $\bA_2$ with probability $1,$ then its variance must be zero. This was the initial point in the EPR critique of quantum mechanics. They argued that in this case, the formalism is self contradictory, because now one could measure $\bB_2$ when the system is in state $\phi_{a_2}$ and then \eqref{HUP} would be contradicted. 

But their argument fails for systems with finitely many levels because of \eqref{T1.1} and \eqref{T1.2}. Note that in this case $\langle\phi_{a_i(1)},\bC_2\phi_{a_i(1)}\rangle=C_2(s-a_1(1),s-a_i(1)),$ the $s-a_i(1)$-th element along the diagonal of $\bC_2.$ That is, if $\bC_2$ has a vanishing diagonal, there will be no contradiction with the uncertainty principle. Since the right hand side of \eqref{HUP} vanishes, it does not matter if  the variance of $\bA_2$ vanishes.

\section{Example}
We present the simplest possible example, consisting of two identical two level systems, whose basis states we label as $|1\rangle$ and $|-1\rangle,$ and the operators $\bA,\bB,\bC$ introduced above are:

\begin{equation}\label{ops}
\bA_j=\begin{pmatrix} 1&0\\0&-1\end{pmatrix},\;\;\;\;\bB_j=\begin{pmatrix}\; 0&1\\1&0\end{pmatrix},\;\;\;\bC_j=\begin{pmatrix}0&-i\\i&0\end{pmatrix}.\;\;\;j=1,2.
\end{equation}
These are the Pauli spin matrices, except for the fact that we chose the units so that $[\bA_j,\bB_k]=2i\bC\delta_{j,k},$ etc. Notice as well that $\bA_j|\pm1\rangle=\pm|\pm1\rangle$ when acting on the corresponding subsystem. Therefore, when considering $\bS=\bA_1+\bA_2$ acting on the composite system, its eigenvalues will be $\{2,0,-2\},$ and the eigenvalue $0$ has a two-dimensional eigenspace.

Suppose that we start from a system prepared in a state $\Psi$ (or $|\Psi\rangle$ when typographically convenient) specified as 
$$\Psi=a(1,1)|1,1\rangle+a(1,-1)|1,-1\rangle+a(-1,1)|-1,1\rangle+a(-1,-1)|-1,-1\rangle.$$
To prepare the system in an eigenstate of given total value of $\bS$ we form:

\begin{equation}\label{projs}
\Pi_2 =|1,1\rangle\langle 1,1|,\;\;\;\Pi_0=|1,-1\rangle\langle 1,-1|+|-1,1\rangle\langle -1,1|, \;\;\;\Pi_{-2}=|-1,-1\rangle\langle -1,-1.
\end{equation}

\begin{gather}
\Psi_2 =\frac{\Pi_2\Psi\rangle}{\sqrt{tr(\Pi_2|\Psi\rangle\langle\Psi|)}}=\frac{a(1,1)}{|a(1,1)|}|1,1\rangle. \;\;\label{eig2}\\
\Psi_0 =\frac{\Pi_0\Psi}{\sqrt{tr(\Pi_0|\Psi\rangle\langle\Psi|)}}=\frac{a(1,-1)|1,-1\rangle+a(-1,1)|-1,1\rangle}{\sqrt{|a(1,-1)|^2+|a(-1,1)|^2}}\;\;\nonumber\\
=\gamma(1,-1)|1,-1\rangle+\gamma(-1,1)|-1,1\rangle.\; \;\;\label{eig0}\\
\Psi_{-2}=\frac{\Pi_{-2}\Psi\rangle}{\sqrt{tr(\Pi_{-2}|\Psi\rangle\langle\Psi|)}}=\frac{a(-1,-1)}{|a(1,1)|}|-1,-1\rangle. \;\;\label{eig-2}
\end{gather}

The identifications in the second identity should be clear. Each of the eigenvalues (and their corresponding eigenstates) will occur with probability 
$|a(1,1)|^2,$ $\,|a(1,-1)|^2+|a(-1,1)|^2$ and $|a(1,1)|^2$ respectively. The analogue of the case considered by EPR corresponds to the value $\bS=0.$ Let us examine what happens to the predictions of $\bA_i$ and those of $\bB_i$ in the state $\Psi_0.$ 

A simple computation yields:
\begin{equation}\label{obs1}
\langle\Psi_0,\bA_1\Psi_0\rangle =|\gamma(1,-1)|^2-|\gamma(1,-1)|^2\equiv \alpha_1
\end{equation}
Clearly, $-1\leq\alpha_1\leq 1.$ At this point we stress the difference between experiment, measurement and prediction. An experiment is a sequence of measurements made on a collection of identically prepared systems. In our case, the measurements of $\bA_1$ on $\Psi_0$ yield values $+1$ and $-1,$ with probabilities (asymptotically) equal to $|\gamma(1,-1)|^2$ and $|\gamma(-1,1)|^2$ respectively. The number in \eqref{obs1} is what the experimenter reports after averaging over all measurements. This number is what theory predicts if the state $\Psi_0$ is known and  $\langle\Psi_0,\bA_1\Psi_0\rangle$ is computed instead of being measured, and the quantity reported in \eqref{obs2} is to be thought as the experimental error or the prediction error depending on the interpretation of \eqref{obs1}. 

Since $\bA_1^2=\bbI,$ and therefore $(\bA_1-\alpha_1\bbI)^2=(\bbI-2\alpha_1\bA_1+\alpha_1^2)$ we conclude that:
\begin{equation}\label{obs2}
\langle\Psi_0,(\bA_1-\alpha_1\bbI)^2\Psi_0\rangle=(1-\alpha_1^2)\;\;\Leftrightarrow\;\;\Delta_{\Psi_0}(\bA_1)=\sqrt{(1-\alpha_1^2)}.
\end{equation}
A similar computation yields the following result, which also follows from Theorem
 \ref{T2}, and is intuitive because of  the symmetry with respect to the exchange of the particles, and the fact that $\langle\Psi_0,\bS\Psi_0\rangle=0$:
$$\begin{aligned}
\langle\Psi_0,\bA_2\Psi_0\rangle =-\langle\Psi_0,\bA_1\Psi_0\rangle =-\alpha_1.\\
\Delta_{\Psi_0}(\bA_2) = \sqrt{(1-\alpha_1^2)}.
\end{aligned}$$
It is also simple to verify that:
\begin{equation}\label{obs3}
\langle\Psi_0,\bB_1\Psi_0\rangle= \langle\Psi_0,\bB_2\Psi_0\rangle = 0.
\end{equation}
Again, since $\bB_i=I,$ using \eqref{obs3} we obtain that:
\begin{equation}\label{obs4}
\langle\Psi_0,\bB_i^2\Psi_0\rangle=1\;\;\Leftrightarrow\;\;\Delta_{\Psi_0}(\bB_i)=1.
\end{equation}
And to finish, another simple computation yields that:
\begin{equation}\label{obs5}
\langle\Psi_0,\bC_1\Psi_0\rangle= \langle\Psi_0,\bC_2\Psi_0\rangle = 0.
\end{equation}
Therefore we have:
\begin{equation}\label{obs6}
\Delta_{\Psi_0}(\bA_i)\Delta_{\Psi_0}(\bB_i)=\sqrt{(1-\alpha_1^2)}>\frac{1}{2}|tr(\bC_i|\Psi_0\rangle\langle\Psi_0|)|=0.
\end{equation}
Therefore,  there is no contradiction to the uncertainty principle \eqref{HUP} after measuring the total momentum and making predictions with the post measurement state.. 

To continue, suppose that we decide to measure $\bA_1$ in the state $\Psi_0.$  The theory asserts that we  would observe that $+1$ happens with probability $|\gamma(1,-1)|^2$ and $-1$ with probability $|\gamma(-1,1)|^2,$ both introduced in \eqref{eig0}.  Anyway, if the measurement yields $+1,$ we are left with the system in the post-measurement state:
\begin{equation}\label{obs7}
\phi_1=\frac{\Pi_1\Psi_0}{tr(\Pi_1|\Psi_0\rangle\langle\Psi_0|)}=\frac{\gamma(1,-1)}{|\gamma(1,-1)|}|1,-1\rangle.
\end{equation}
Note that $\bA_1\phi_1=\phi_1$ and $\bA_2\phi_1=-\phi_1.$ Not only that, $\Delta_{\phi_1}(\bA_i)=0$ for $i=1,2.$ As noted in Theorem \ref{T1}, in this case $\langle\phi_1,\bC_i\phi_1\rangle=0,$ and thus no contradiction with the uncertainty relation \eqref{HUP} arises. Nevertheless, in this example $\langle\phi_1,\bB_i\phi_1\rangle=0$ as well. But since \eqref{obs5} is in force, the objections raised by EPR are not valid in this case, and there are no spooky actions at a distance.

\section{Quantum conditional expectations}
The notion of quantum conditional expectation is quite similar to the classical notion of conditional expectation, except that the technicalities are much larger. For the application that we need, since the operators involved ($\bA_1,$ $\bA_2$ and $\bS$) commute, the technicalities disappear, and the definitions coincide with those of the classical case.  Next we collect two basic properties and establish the correspondence with the probability distributions in the post-measurement state. To use a suggestive notation, let $\bO$ be a Hermitian matrix denoting  an observable of the system and put as in \cite{Gu2}:
\begin{equation}\label{exp}
\cE_\Psi[\bO] = tr(\bO|\Psi\rangle\langle\Psi|) = \langle\Psi|\bO|\Psi\rangle.
\end{equation}
Now we have

\begin{definition}\label{CE0}
Suppose that our system is in a state $\Psi\in\cH$ and let $f(\bA_1)$  The quantum conditional expectation of  $f(\bA_1)$   given an observation of  $\bS,$ is an operator valued function of $\bS,$ denoted by  $\cE_{\Psi}[f(\bA_1)|\bS]$ that satisfies
\begin{equation}\label{CE1}
\cE_{\Psi}\big[\cE_\Psi[f(\bA_1)|\bS]G(\bS)]\big] = \cE_\Psi[f(\bA_1)G(\bS)],
\end{equation}
for any $G(\bS).$
Furthermore, the conditional expectation operator satisfies:
\begin{equation}\label{CE2}
\cE_\Psi[H(\bA_1,\bS)|\bA_1,\bS]= H(\bA_1,\bS)
\end{equation}
for any $H(\bA_1,\bS).$
\end{definition}

The notation used in classical probability to  write \eqref{CE1} is $\cE_{\Psi}[f(\bA_1)|\bS=p]$ which is to be understood as the value of the function $\cE_{\Psi}[f(\bA_1)|\bS]$ when $\bS$ is observed to have value $s$ in a measurement. Similarly, in \eqref{CE2}, when an observation of  $\bS$ and $\bA_1$ yields values $s$ and $a_1$ respectively, then 
\begin{equation}\label{interp1}
\cE_\Psi[H(\bA_1,\bS)|\bA_1=s,\bS=s]= H(a_1,s).
\end{equation}

Let us denote the value of the function $\cE_{\Psi}\big[\cE_\Psi[f(\bA_1)|\bS]$ when $\bS=s$ by $e(s_k).$ To obtain the quantum conditional probability density using \eqref{CE1} we compute both sides of it, use the fact that 
$G(\bS)$ is an arbitrary function of $\bS,$ and then solve for the quantum conditional expectation.

$$\begin{aligned}
&\cE_{\Psi}\big[\cE_\Psi[f(\bA_1)|\bS]G(\bS)]\big]=\sum_{a_i,a_j} e(a_i+a_j)G(a_i+a_j)q_\Psi(a_i,a_j)\\=
\sum_{k=1}^D&\sum_{a_i\in\cS(k)}e(s_k)G(s_k)q_\Psi(a_i,s_k-a_i)=\sum_{k=1}^DG(s_k)e(s_k)\sum_{a_i\in\cS(k)}q_\Psi(a_i,s_k-a_i)\\
\sum_{k=1}^DG(s_k)e(s_k)p(s_k).
 \end{aligned}$$ 
A similar computation for the right hand side goes as follows:
$$\begin{aligned}
&\cE_\Psi[f(\bA_1)G(\bS)] = \sum_{a_i,a_j} f(a_i+a_j)G(a_i+a_j)q_\Psi(a_i,a_j) \\
\sum_{k=1}^D&\sum_{a_i\in\cS(k)}f((a_i)G(s_k)q_\Psi(a_i,s_k-a_i)=\sum_{k=1}^DG(s_k)\sum_{a_i\in\cS(k)}f(a_i)q_\Psi(a_i,s_k-a_i)\\
 &\sum_{k=1}^DG(s_k)p(s_k)\sum_{a_i\in\cS(k)}f(a_i)\frac{q_\Psi(a_i,s_k-a_i)}{p(s_k)}. 
\end{aligned}
$$
Therefore:
$$ \cE_\Psi[f(\bA_1)|\bS=s_k]=e(s_k)=\sum_{a_i\in\cS(k)}f(a_i)\frac{q_\Psi(a_i,s_k-a_i)}{p(s_k)}$$

In other words, the expected value of $f(\bA_1)$ in the post measurement state as in \eqref{pred3} or \eqref{rs2}, and its quantum conditional expectation given $\bS$  as in \eqref{CE1} coincide. 

\section{Concluding remarks}
The discrete case is different from the continuous case in one essential detail. According to \eqref{T1.2} the right hand side of the uncertainty principle \eqref{HUP} vanishes and there is no problem if $\Delta_\Psi(\bA_1)=0$ and $\Delta_\Psi(\bA_1)\not=0.$

In the continuous case, If $\bA$ is the momentum operator and $\bB$ is the position operator, then $\bC$ is a multiple of the identity. In this case the argument is more elaborate and one of the variances must be infinite when the other vanishes. This case was analyzed in \cite{Gz}.

The other issue worth mentioning is that the predictions after a measurement is made, are conditional predictions (predictions made with conditional probabilities), and if we were interested in unconditional predictions, we have to average with respect to the original probability distribution of the measured variable.

\section{Declaration: Conflicts of interest/Competing Interests}
I certify that there is no actual or potential conflict of interest in relation to this article.


\begin{thebibliography}{}
\bibitem{A}  Aspect, A and Grangier, P. (2005). {\it De l'article d'Einstein Podolsky et Rosen \'a l'information quantique: les stup\'efiantes propri\'et\'es de l'intrication}, Chapter 2 in Einstein Aujourd'hui, (Leduc, M and Le Bellac, N, Eds), De Gruyter Brill,  EDP Sciences/CNRS Editions, Paris.

\bibitem{EPR} Einstein, A., Podolsky, B. and Rosen, N. (1935). {\it Can quantum mechanical description of reality be considered complete}, Physical Review, {\bf 47}, 777-780. \url{https://doi.org/10.1103/PhysRev.47.777}

\bibitem{Gu1} Gudder, S. (1979) . {\it Stochastic Methods in Quantum Mechanics}, Elsevier North Holland Inc., New York.

\bibitem{Gu2} Gudder, S. (2024). {\it Quantum transition probabilities}. Available at \url{https://arxiv.org/pdf/2404.00177}.

\bibitem{Gz} Gzyl, H. (2024). {\it A predictive solution to the EPR paradox}, (To be submitted elsewhere), \url{http://arxiv.org/abs/2508.20788}.

\bibitem{He} Helstrom, C.W. (1961). {\it Quantum Detection and Estimation Theory}, Academic Press, N.Y.

\bibitem{Ho} Holevo, A.S. (1982). {\it Probabilistic and Statistical Aspects of Quantum Theory}, North Holland Publishing Company, Amsterdam.

\bibitem{J} Jacobs, K. (2014). {\it Quantum Measurement Theory}, Cambridge University Press, Cambridge.

\bibitem{Ku} Kupczynski, M. {\it Seventy Years of the EPR Paradox},  \url{ 	
https://doi.org/10.48550/arXiv.0710.3397}

\bibitem{LeB} Le Bellac, M. (2013). {\it Physique Quantique}, EDP Sciences/CNRS Editions, Paris.

\bibitem{La} Lalo\"{e}, F. (2019). {\it Do We Really Understand Quantum Mechanics, 2nd Ed.}, Cambridge University press,  Cambridge.

\bibitem{www} \url{https://scispace.com/topics/epr-paradox-1lcbi0vy}

\end{thebibliography}
\end{document}